\documentstyle[12pt,aps]{revtex}
\def\Sa{S^\alpha}
\def\a{\alpha}
\def\one{$1/f$}
\title{Dynamical systems theory for music dynamics}
\author{Jean Pierre Boon \footnote{email address: {\tt jpboon@ulb.ac.be}}
        and Olivier Decroly\\
        Physique Nonlin\'eaire et M\'ecanique Statistique \\
        Universit\'e libre de Bruxelles, Campus Plaine CP 231\\
        B-1050 Bruxelles, Belgium}
\date{{\it to appear in} {\bf CHAOS}}
\begin{document}
\maketitle
\begin{abstract}

 {We show that, when music pieces are cast in the form of time series of
 pitch variations, the concepts and tools of dynamical systems theory can be
 applied to the analysis of {\it temporal dynamics} in music. (i) Phase space
 portraits are constructed from the time series wherefrom the dimensionality
 is evaluated as a measure of the {\pit global} dynamics of each piece.
 (ii) Spectral analysis of the time series yields power spectra
 ($\sim f^{-\nu}$) close to {\pit red noise} ($\nu \sim 2$) in the low
 frequency range.
 (iii) We define an information entropy which provides a measure of the
 {\pit local} dynamics in the musical piece; the entropy can be interpreted
 as an evaluation of the degree of {\it complexity} in the music, but there
 is no evidence of an analytical relation between local and global dynamics.
 These findings are based
 on computations performed on eighty sequences sampled in the music
 literature from the 18th to the 20th~century.}
\end{abstract}
olatex  musalm.tex
\section{Quantitative approach to music analysis}\label{1}

   Any attempt to investigate the forms of artistic expression through
quantitative analysis calls necessarily, in one way or another, for the use of
methods developed with scientific techniques, and thereby encounters
unavoidably the difficulties inherent to the quantification of art.  Certainly
music is the art form which by construction should be best amenable to
scientific approach: (i) western music proceeds by vertical (harmony) and
horizontal (counterpoint) structures; (ii) its language and syntax are based on
mathematical properties where symmetries and symmetry breaking play an
essential role; (iii) music, as performed, steers a one-dimensional course, the
dimension of time.  So time irreversibility is intrinsic to musical expression
and a musical sequence can be considered as the time evolution of an acoustic
signal.  The signal detection (via human acoustical perception or via technical
audio-equipment) triggers the sequence: measurement - coding - analysis,
followed eventually by interpretation.  As complicated as they may be
(physiologically or technically), measurement and coding are essentially
operational steps.  Analysis is truly the key-point: it is a conceptual process
whose results are indispensable to meaningful interpretation.

Starting from the idea that we receive musical messages as the time
evolution of acoustic signals, we can associate to the development of
the musical sequence the concept of time series and use the analytical
and computational tools of physics and mathematics for the analysis of
{\it music dynamics} \cite{fo0} . Previous work along these lines
\cite{Vos,Boo,Net} gave indication that time correlations and spectral
characteristics could be identified in pieces of music chosen as typical
examples. So the feasibility of the scientific approach to music analysis has
been established, but clearly the results are limited in number and far from
conclusive in scope.  In particular they call for a more thorough investigation
of a large number of pieces covering the whole history of music.  In the next
section we review earlier work by way of introduction to, and justification of
the present work, whose methods are developed in section \ref{3}.  These
methods
were applied to the analysis of about eighty musical sequences and the results
are discussed in section \ref{4}.  We conclude with some comments.

\section{\one \, music and dimensionality}\label{2}

	Obviously there is more than one single time signature to a musical work:
whether one considers the first successive few notes of a piece (say Bach's
third {\it Brandenburg Concerto}), an entire movement, the full piece, or the
complete set of such pieces, one finds different time characteristics.
Schematically however two time scales emerge: a short time scale and a long
time scale.  The {\it short time} scale characterizes the dynamics of a musical
cell, i.e. a group of ten to twenty successive notes or sounds which are highly
correlated in time.  Such short time correlations are found in almost any music
(unless constructed on the basis of a white noise algorithm) and are therefore
not very useful for a quantitative analysis designed to characterize works,
composers or styles.
	{\it Long time} correlations have been investigated by Voss and
Clarke \cite{Vos}
 by spectral analysis of the audio signal of recorded music.  The
analysis proceeds with complete pieces of music (e.g. Bach's
first {\it Brandenburg Concerto}) as well as with up to twelve hour stretches
of data accumulated from radio stations.  Low pass filtering was performed
to obtain spectra of these data in the low frequency domain.
Voss and Clarke found that loudness variations and pitch
fluctuations exhibit $1/f$ power spectra in the low frequency range
($f\leq 10 Hz$)
independently of the music considered.  This finding raised interest and
stimulated further work amongst musicians and scientists (see \cite{Net}
for references).  The claim that frequency fluctuations in music have a $1/f$
spectral density has recently been critically discussed by Nettheim \cite{Net}
in a constructive analysis.  In particular Nettheim questions the validity of
the procedure when long duration stretches include various pieces, different
composers and styles as well as spoken announcements and comments.  He
emphasizes that "a single piece is normally the largest unit of artistic
significance" (\cite{Net},  p.136), a statement commonly endorsed by
musicians.  Nettheim provides new data for single movements of 18th-19th
century classical music
 yielding spectral analysis results which are at variance with
$1/f$ behavior, while he recognizes that it would be desirable to extend the
analysis beyond the five examples treated in his study.

 A different approach towards a quantitative analysis of music has been
proposed by Boon {\it et al.} \cite{Boo} who showed that the theory of
dynamical
systems could provide interesting tools for the identification of complex
dynamics in music.  The basic material used by Boon {\it et al.} is the printed
score
played on a synthesizer interfaced to a computer; pitch values are converted
into digital data which are stored in the computer memory so that the score
is converted into a time series, say $X(t)$ for a single part score
(see Fig.1a). Pieces with several parts are treated part by part to produce
a set of time series $X(t), Y(t), Z(t),...$ .  Data processing is performed
to construct the phase portrait: $X(t), Y(t), Z(t),...$ or
$X(t), X(t+n\Delta t), X(t+m\Delta t)$, using the time-delay method for
single part pieces.  The phase portraits (see Fig.1b) are used to compute
the dimensionality, and the initial time series to calculate the correlation
function $C(t)$ (see Fig.1c) and the corresponding power spectrum $S(f)$
(see Fig.1d) \cite{Boo}. Here we also consider the information
entropy which is discussed in section \ref{3}.

 The phase portrait constitutes a spatial
representation (in the abstract phase space) of the temporal dynamics of the
music piece reconstructed from the time series obtained from the pitch
variations as a function of time.  An example is given in Fig.1b which shows
the phase portrait of the three part {\it Ricercar} of Bach's {\it Musical
Offering}.  The
Hausdorff dimension is computed with the box-counting method.  As references
we consider two opposite cases.  (i) A piece of random music constructed with
a computer generated white noise algorithm exhibits a phase space trajectory
which fills homogeneously the entire phase space and consequently has
dimensionality three in a three-dimensional phase space.  (ii) An elementary
score constructed as a canon of three repeatedly ascending and descending
chromatic scales yields a limit cycle with dimensionality one.  For the six
pieces of music analyzed by Boon {\it et al.} , the Hausdorff dimension takes
values around two, a result which calls for further analysis in view of the
limited number of data.

The power spectrum of the time series obtained from the
scores was also considered by Boon {\it et al.} \cite{Boo}; for instance
the $log-log$
 plot of  $S(f)$ of the first part of Bach's {\it Ricercar} shown in Fig.1d
exhibits, in the range $0.03 - 3.0 Hz$, an average slope
corresponding to $1/f^\nu$ with $1<\nu<2$.
However no systematic spectral analysis of the pieces studied was presented.

\section{Method and analysis}\label{3}

The pieces to be analyzed are played by a musician who uses a synthesizer
interfaced to a computer where the pieces are stored.  At this stage the
computer is used as a multitrack taperecorder: musical sequences can be played
at will independently or simultaneously in any combination and quantitative
corrections can be made to the recorded tracks to obtain perfect match  with
the original score. The pieces are stored in files which can be called for
treatment and analysis. Digitizing the musical pieces requires double
discretization, i.e. for pitch and for duration \cite {fo1}.
 The time scale is
set by the shortest time step required to account for the shortest duration
used in the piece, typically the duration of one note of the 64th note triplet,
which corresponds to $2/100$ sec. for a tempo set as quarter note $120$.
For the pitch scale we use the well-tempered clavier, i.e. the half-tone is
the basic unit.

The choice of the music pieces was guided by several criteria.
Since we are using the musical sequence as a time series, the pieces should
ideally have parts with successions of exclusively
single notes, like e.g. in woodwind chamber music.  String trios, quartets,...
fulfill this criterion if one accepts to perform a reduction of occasional
double strings.  Counterpoint pieces (e.g. fugues) are amenable to analysis
after proper decryption of the individual parts.  Jazz music as written on
transcribed scores can be cast, for the purpose of the present investigation,
into three parts: the melody, the bass line and the middle part (although
such a drastic reduction certainly overlooks the essentials of jazz).

Another constraint concerns the length of the pieces to be analyzed: the number
of notes must be sufficiently large (in practice not smaller than 500) to yield
acceptable data for time series analysis.  In addition to these technical
criteria, we wanted a sample of pieces covering significantly the history of
western classical music.  Twenty-three pieces were chosen, which, accounting
for each part in each piece, amounted to
the coding of eighty-three sequences, from J.S. Bach to E. Carter, plus four
jazz music scores.  The complete list is given in Table~1.

The method was tested
against two "idealized" sequences whose analysis provided reference values for
the music pieces data.  The first sequence constructed from repeated ascending
and descending scales serves a typical example of deterministic (periodic)
dynamics.  On the other hand, we generated a sequence of random
music (based on a white noise algorithm) as a prototype of unpredictable music.
As the data to be obtained from the analysis of these sequences can be
predicted theoretically, these reference pieces also serve the purpose to
evaluate quantitatively the influence of the number of notes in a sequence on
the value of the measured quantities.

 We developed and used three methods for
the analysis of temporal dynamics in music: (i) the phase space portrait and
its quantification through dimensionality ; (ii) the autocorrelation
function of the time series and its power spectrum; and (iii) the entropy.

(i) The phase portrait method which maps the time evolution of a dynamical
process onto a spatial representation is straightforwardly borrowed from
dynamical systems theory and its application to music dynamics was introduced
by Boon {\it et al.} in \cite{Boo}. We call attention to the fact that,
since pitch and time variables are discretized (see above), the phase space
is discrete and
finite as it is spanned by the dynamical variables $X(t), Y(t), Z(t),...$
corresponding to the time variations of the pitch in each part of the music
score.  The size of the phase space is set by the largest pitch extent in the
piece.  An example is shown in Fig.1b.

	Since the phase space is discrete for musical trajectories,
the Hausdorff dimension

\begin{equation}
D_H = \lim_{\lambda \rightarrow 0} \log {\cal N}(\lambda)\Big/ \log (1/\lambda)
\end{equation}
 must be redefined appropriately because the limit $\lambda \rightarrow 0$
 is immaterial
here.  We denote by $\mu$ the size of the smallest box that can be constructed,
i.e. $\mu$ is the half-tone unit.Then

\begin{equation}
 {\cal N}(\lambda)\times (\lambda /\mu)^{D_f} = V/\mu ^{D_f}
\end{equation}
where the {\it rhs} is a constant.  It follows that

\begin{equation}
 D_f = [-\log {\cal N}(\lambda ) + const.]\Big / \log (\lambda /\mu)
\end{equation}
 and the dimensionality is obtained as the slope of
 $log \, {\cal N}(\lambda )$  versus $log \, \lambda$  :

\begin{equation}
\log {\cal N}(\lambda) = -D_f \log\lambda + const.' \quad;\quad (\lambda \ne
0).
\end{equation}
where $log \, \mu$ has been absorbed in the constant.

 The dimension $D_f$ is computed by
the box-counting method and, as it characterizes the structure of the complete
phase trajectory, its value yields a quantitative evaluation of the global
dynamics of the musical piece.

(ii) Power spectrum methods are well known and need not be described here.
$Log-log$ plots of the spectra obtained from the
time series of the musical sequences are of interest in that the trend of
their slopes yields a value which has been claimed to be "universally"
typical of $1/f$ noise, a claim which has been critically discussed
subsequently (see section \ref{2}). In the next section we present the
results obtained for the pieces analyzed in the present work.

(iii) While the Hausdorff dimension provides an
evaluation of the {\it global} dynamics of a piece of music, further
information on the {\it local} dynamics can be obtained from the application
of information theory. The analysis proceeds on the basis of the data files.
 A sequence of notes can be viewed as a string of
characters and as such can be analyzed from the point of view of its
information content.  The simplest string follows from straightforward
coding of the pitch by assigning a character to each note.
Duration of notes is included by repeating the character as many times as
there are unit time steps until the next note (rests can be accounted for
by incorporating a specific symbol for rest unit). Sequences
of pitch intervals can be coded similarly with a signed number corresponding to
the number of half-notes up or down from one note to the next.  (Note that the
interval coding is independent of the key).

The entropy which measures the
information content of a string of characters on the basis of their occurrence
probability, will be defined such that its value has an upper bound ($=1$)
for a fully random sequence. $H$ denotes the entropy when strings of notes are
considered independently of their duration; when the latter is accounted for,
we use $H^r$; $H^i$ is used for strings of intervals.  Furthermore, a
subscript
indicates the order: $H_0$ is the zeroth order entropy which is a measure of
the
straight occurrence of each note, and the $\alpha$-th order entropies,
$H_\alpha$ ($\alpha \neq 0$), follow from the successive conditional
probabilities at increasing orders.  We now formalize these quantities.

Consider a string of characters $S = \{s_i; i=1,..., N\}$ where the $s_i$'s
can be chosen from an alphabet with $R$ characters.
Subsequently the $N$ characters will be identified with the notes of a musical
sequence - as its symbolic dynamics - and the alphabet will be identified
with the degrees in the pitch range over which the piece of music extends.
The set $\{S^{\alpha} _j; j=1,...,n;n=N-\alpha + 1\}$  of strings
containing $\alpha $ characters ($\alpha \leq N$) is obtained by partitioning
the initial set as follows

\begin{equation}
\{\Sa_1=(s_1,...,s_\a),...,\Sa_j=(s_j,...,s_{j+\a -1}),...,\Sa_n=
 (s_{N-\a +1},...,s_N)\}
\end{equation}
The occurrence probability of a given string $\Sa$ is defined by

\begin{equation}
 P(\Sa)=\nu^\a/n_\a ,
\end{equation}
where $\nu^\a(=\nu(\Sa))$ is the number of occurrences of string
$\Sa$ and $n_\a$
is the total number of substrings of $S$ with $\a$ characters.  We now
consider substrings with $(\a+1)$ characters where the first $\a$
characters belong to a given string $\Sa$ and we define the probability

\begin{equation}
 P(s\mid \Sa)=\nu_{s\mid \a}\,\Big/\,\sum_{s'\in R}\nu_{s'\mid \a}
\end{equation}
as the relative measure of the number of occurrences $ \nu_{s\mid \a} $
of character $s$ given that the $\a$  previous characters belong to the
string $\Sa$. The corresponding normalized entropy is given by

\begin{equation}
 H(\Sa)=-\sum_{s\in R}P(s\mid\Sa)\log\,P(s\mid\Sa)\,\Big/\,\log \nu^\a.
\end{equation}
 Then performing the sum of the entropies corresponding to all possible
 strings with $\a$ characters, weighted by the occurrence probability of
each  $\Sa$ yields the $\a$-th order entropy

\begin{equation}
 H_\a=\sum_{\Sa} H(\Sa) P(\Sa)
\end{equation}
{}From (4) it is clear that $0\leq H(\Sa) \leq 1$, and since by definition
   $\sum_{\Sa} P(\Sa) =1$ (see(6)), it follows that $0\leq H_\a \leq 1$.
We now comment on the meaning
of these quantities in the context of musical sequences.  The meaning of
$H_0$ is straightforward (see comment above), but not very useful: $H_0=0$
when the score uses a single note and $H_0=1$ when all accessible degrees
are equally visited (a chromatic scale or a random score). $H_1$ is related
to the probability to find the note $s_{i+1}$ given that the previous note
was $s'_i$: $H_1$ has the value one for a random score as well as for a
descending and ascending scale. $H_2$ is a more interesting quantity in
that it discriminates between deterministic sequences and random sequences:
obviously $H_2=1$ for a random score and $H_2=0$ for a scale.  Higher order
entropies can be considered, but we found that they take fast decreasing
values as the length of the string increases.

 When these concepts were
applied to musical sequences, it appeared in the course of the analysis,
that the numerical results did not reflect consistent significance.  The
entropies are measures of the information content of the sequences and
thereby provide an evaluation of the "diversity" of the notes in the score.
 However as most pieces analyzed are tonal music, there is one important
feature which must be accounted for: {\it tonality}.  We now introduce a new
entropy which incorporates the property that the notes in a sequence
 belong or not to a reference scale (e.g. given by the clef, but not
necessarily).

	A scale is defined by a tonality (A, B$\flat$,...) and a mode
(Major, minor,...); the degrees of the scale, starting with the tonic,
yield a succession of well-defined degrees. $\theta$ is the set
of notes containing  all such degrees. In order to better quantify
the "diversity" of the notes in a musical sequence and thereby attempting
to take into account the "liberty" taken by the composer with tonality, we
discriminate notes which belong to $\theta$ from those which do not.
Mathematically, this is accomplished by separating the probability $P_s$
of occurrence of note $s$  into two contributions ($P_s'$) depending on
whether $s$ belongs to $\theta$ or not. $P_s$ is the straight occurrence
probability $P_s= n_s/\sum^R_{s'=1}n_{s'}$ , where $n_s$ is the
number of occurrences of note $s$, and the denominator is the total number of
notes ($N$) in the sequence.  Then we define

\begin{equation}
 P_s'=\left\{ \begin{array}{ll}
               n_s/\sum_{s\in \theta} n_s &\mbox{if $s\in \theta$;}\\
               n_s/\sum_{s\not\in \theta} n_s &\mbox{if $s\not\in\theta$.}
              \end{array}
      \right.
\end{equation}
We now associate different weighting factors, $\gamma$ and $\delta$,
to occurrences within and out of $\theta$ respectively, such that

\begin{equation}
 \gamma \sum_{s\in \theta} P_s' + \delta \sum_{s\not\in \theta} P_s' =1
\end{equation}
with $\gamma + \delta =1$. As $\gamma$ and $\delta$ are treated as
parameters, we define the new normalized entropy

\begin{equation}
 H'=-[\sum_{s\in \theta} \gamma P_s' \log(\gamma P_s') +
\sum_{s\not\in \theta} \delta P_s' \log(\delta P_s')]
 \Big/\log N              \end{equation}
as the {\it parametric} entropy.  The idea being that deviations from
tonality are considered as "unexpected events", thereby contributing more
 strongly to an increase of entropy, the values of $\gamma$ and $\delta$
will be set (empirically; see section \ref {4}) such that $\delta=m\gamma$
with $m > 1$~: occurrences of notes off-tonality ($s\not\in \theta$) are given
a more important weight than those of notes belonging to the reference scale.
 So $H'$ should be low for pieces of music where most notes are within a
well-set tonality, and should be high for nontonal music.  Furthermore, for a
strictly twelve-tone piece the value of $H'$ should be independent of any
reference key.  On the other hand we set $\gamma =1$ and $\delta =0$ for a
sequence where all $s\in \theta$.
  For random music, the parametric entropy is independent of $\gamma$
and $\delta$ and its
value approaches one when the number of notes in the sequence becomes
sufficiently large.

 Parametric entropies can also be defined to successive orders

\begin{equation}
 H_\a '= \sum_{\Sa} H'(\Sa) {\cal P}(\Sa)
\end{equation}
where

%\begin{equation}
% H'(\Sa)  = -\sum_{\eta =\gamma ,\delta ;} \sum_{i=\a +1}^N \eta
% P(i \mid \Sa_k)
% log  \lbrack \eta P(i \mid \Sa_k) \rbrack /log\nu^\a_k
%\end{equation}
\begin{eqnarray}
H'(\Sa)\,=\,-&\Big[\sum_{s(\in \theta)}&
\gamma P(s \mid \Sa)
\log (\gamma\,P(s \mid \Sa)\,)\,+  \nonumber \\[12pt]
      &\sum_{s(\not\in \theta)}& \delta P(s \mid \Sa)
\log (\,\delta P(s \mid \Sa)\,)\Big]\Big/(\log\,\nu^\a)
\end{eqnarray}
%with  $i \in \theta$ for $\eta = \gamma$ and $i \not\in \theta$
%for $\eta = \delta$,
and $ {\cal P}(\Sa)$ is the weighted probability of occurrence of string
$\Sa$, i.e. if this string contains $\kappa$ characters $\in \theta$
(and $(\a -\kappa) \not\in \theta$)

\begin{equation}
 {\cal P}(\Sa) = \gamma^\kappa \delta^{\a-\kappa}\nu^\a/n_\a.
 \end{equation}
To first order the interpretation is as follows: $H'_1$ yields a measure of
the information content of a musical sequence quantifying the transition
probabilities from one note to the next given that the transition can occur
within the reference tonality ($(s_i,s_{i+1}) \in \theta$), outside the
tonality
($(s_i,s_{i+1}) \not \in \theta$), or from $\theta$ to off  $\theta$
($s_i\in \theta,s_{i+1}\not \in \theta$), and vice-versa.
The operational result is that a large value of the parametric
entropy is indicative of frequent excursions away from the tonality, with
transitions over intervals distributed over a large number of notes. On
the contrary, the entropy will assume a low value when a note determines
almost unambiguously the next one, in particular when the next note remains
in the range of tonality.
  Higher order parametric entropies are neither obviously
interpreted nor very useful in practice: their values decrease
uniformly very rapidly (because of the factors
$\gamma^\kappa \delta^{\a-\kappa}$ in (11)) and yield
no significantly distinguishable data for the pieces analyzed.

\section{Results and comments}\label{4}

A list of the pieces of music which were analyzed is given in Table 1.
The analysis was performed for each part of each piece; thirteen
 quantities were computed for eighty-three sequences yielding a large
number of numerical data wherefrom a selection was made by significant
criteria analysis.  The selected results are shown in Table 2.  The Hausdorff
dimension $D_f$ (whose value is given in the first column of Table 2) is the
mean value of the dimensions computed for each part by the time delay method
($D_f^R$).
The reason for using this quantity (rather than the "global" dimensionality
obtained directly from the phase portrait in $D$-dimensional space)
is that the other quantities (entropies, spectral characteristics) are
computed for each part and so must be averaged
to characterize the complete piece.  Furthermore we obtain good
agreement between the value of the "global" dimensionality and those of
the reconstructed trajectories (a property which is consistent with
 Ruelle-Takens theorem, despite the fact that here the phase space is
 discrete).

	The computation of the parametric entropy was performed not only with
the tonality given by the clef as the reference tonality but also with
 respect to other tonalities.  It would be expected that, as a general
 rule, the minimum value of the entropy should be obtained for the tonality
corresponding to the clef.  We found that this is not necessarily the case.
The values given in Table 2 are the minimum entropies obtained and
they may as well correspond to a neighboring tonality or to the
corresponding major tonality for a piece written in minor tonality.
 Optimization criteria were also used to set the
values of the parameters $\gamma (=0.2)$  and $\delta (=0.8)$.

	An interesting result is
 obtained by considering the Hausdorff dimension (of the reconstructed
trajectories) as a function of the first order parametric entropy: figure 2
shows $log D_f^R = f(H'_1)$. While there is no evidence of
 quantitative correlation between the global dynamical structure (as
measured by $D_f^R$) and local dynamics (the information content evaluated by
$ H'_1$) , it is interesting that about all data are clustered within a
limited zone of the plane $(D_f^R, H'_1)$. In figs.3 and 4 the values of
 $D_f^R$ and $H'_1$ respectively are
 presented in chronological order.  Note that the value $D_f^R = 3$ for
the randomly generated sequence is obtained when the number of data is large
 ($5 \times 10^3$ notes).  Since none of the musical sequences contains
 as many notes, we show the values of  $D_f^R$ for random
 sequences generated with $500$
 and $2000$ notes for reference.  The important observation here is that -
with very few exceptions -  there is no obvious clustering of pieces by
composer or by period of composition; this holds for dimensionality as well
as for parametric entropy.

	A similar comment can be made when
considering Fig.5.  Here - as in Fig.2 - we plot the dimensionality versus
parametric entropy, but we use the values obtained by averaging over all
 parts in order to characterize the piece globally (see Table 2 for data).
No systematic grouping appears neither by composer nor by style. While
physicists, who are keen to logical rules, may be
 disappointed by the lack of systematic ordering that could be drawn from
these results, musicians may find it comforting that music resists simple
formalization. Nevertheless some general comments are in order.

(i) We investigated systematically the possible relations
 between all measured quantities (entropies and dimensionality; see Table 2).
 All plots produced widely scattered data without any indication of
 obvious correlations, except $D_f = f(H'_1)$. While no analytical relation
 could be conjectured for $D_f = f(H'_1)$ \cite{fo2}, Fig.5 suggests that a
 trend might emerge from a statistical analysis performed on a sufficiently
 large sampling of music pieces. An extension of the analysis should also
 be considered where the present treatment is generalized to {\it harmonic
 dynamics}, that is instead of considering the {\it melodic lines} from each
 part separately, consider the sequence of the {\it vertical structures} of
 the score obtained by constructing the vectors of the simultaneous notes
 at each time step, define the corresponding transition probability matrix
 in terms of interval changes, and generalize accordingly the entropies of
 section \ref{3}, in particular by introducing chordal references to define
 the parametric entropy.

(ii) The last column of Table 2 shows the value of the slope $\nu$
 of the $log-log$ plot of the power spectrum  $S(f) \sim f^{-\nu}$,
obtained by
least-squares fit computation in the range $0.03 Hz<f<3.0 Hz$.  The values
 shown here are obtained by averaging over the value of $\nu$ from each part
 of each piece (there are but very weak differences between the values of
 each separate part).  The results are remarkably consistent:
$1.79\leq \nu \leq 1.97$
and indicate values which are close to the {\it red noise} value.
The data of the spectral analysis given in Table 2 confirm and quantify
the qualitative result of Nettheim based on the analysis of five
melodic lines (Bach, Mozart, Beethoven, Schubert and Chopin) \cite {Net}.
These findings are at variance with Voss and Clarke $1 / f$ claim \cite {Vos}.
It must be re-emphasized that in the present work as well as in Nettheim's,
the musical sequences considered are single pieces taken as "the largest
unit(s) of artistic significance"(\cite {Net}, p.136) whereas Voss and Clarke
considered long stretches of recorded music (see Section \ref{2}). That such
long time sequences yield a $1/f$ spectrum found a plausible explanation
in a theoretical analysis by Klimontovich and Boon \cite {Kli}. However
if the musical dynamics analysis is meant as a procedure to identify and
characterize elements of musical significance, the single piece is the
commonly recognized object to be studied. In this respect the meaning
of long stretches of blended musical pieces is unclear.

Note the value $\nu \approx 0$ for the random sequences ($\ast 500$,
$\ast 2000$, and $\ast 5000$ in Table 2). These sequences were constructed
by distributing, on equal time intervals, pitch values according to
white noise generated data; they may qualify as examples of "white noise
music" as they exhibit a flat power spectrum \cite{fo3}.

(iii) The unexpected elements in a piece of music can be found in the
deviations from established rules and the violation or even the mere
 rejection of such rules.  In the context of classical forms, these
deviations are mostly related to the liberty taken by the composer
 with respect to tonality.
Thus when Leibowitz \cite{Leb}
 considers {\it the complexity of musical language},
he argues that {\it Bach's and Haendel's complex polyphonic style is commonly
opposed to what has been called the homophony of Haydn and Mozart (...).
 According to which criteria does one evaluate simplicity and complexity?
 Only one: the counterpoint} (...).  However, continues Leibowitz, {\it the
counterpoint is hardly the only constituting element in music, and, even
 more, it should be obvious that music can be simple or complex
independently of any notion of counterpoint}.  Leibowitz then considers
the problem
of harmony and so observes that the composer's {\it audacity as well as
harmonic
complexity may and must be evaluated according to further criteria}.  Those
 then invoked concern {\it  the principles of tonality expansion} and here - as
argued on the basis of a few specific examples - { \it Haydn's and Mozart's
works appear more audacious than those of their precursors}.  Obviously
the argument is of considerable importance as it leads Leibowitz to the
concept of {\it increasing complexity which should determine the overall
evolution of musical tradition}.  Considering that entropy provides a
quantitative measure of the
degree of complexity the present results show that complexity - in contrast
to Leibowitz' hypothesis - appears to be characteristic of the composition
rather than of the composer.  Accordingly we find no indication of
a systematic increase in complexity paralleling historically the
evolution of classical music.

\section{acknowledgements}\label{5}
We thank Olivier Tribel and Dean Driebe for a critical reading of the
manuscript and Nigel Nettheim for valuable comments..
We acknowledge support from the Fonds National de la Recherche
Scientifique (FNRS, Belgium) and from the Communaut\'e Fran\c caise de
Belgique.

%\newpage

\newpage

{\Large \bf Tables.}
\bigskip

\noindent TABLE 1 : Data for the pieces analyzed, presented in chronological
          order: Composer, Year of composition, Title, Movement analyzed
          (unless single movement piece), Key (except for Carter's Quartet),
          Measure. The first column gives the acronym for identification (ID)
          in Table 2 and in figs.3 and 4.
 \bigskip

\noindent TABLE 2 : Numerical results for the pieces analyzed.
          Chronological order as in Table 1 where the acronyms are defined
          (chromatic scale = SCA; random sequences are indicated by the
          number of notes: 500, 2000, 5000; $\ast$ indicates a computer
          generated sequence). The second
          column shows the number of notes in each piece. The numerical
          values are averages taken over the values obtained for each part
          of the piece (unless single part piece, BPS). $H'_1$ and $\nu$
          are not relevant quantities for the chromatic scale (SCA).

\newpage

{\Large \bf Figure Captions.}

\bigskip

\noindent FIGURE 1 : {\it Ricercar} of J.S.Bach's {\it Musical Offering}.

\noindent (a) Time series: Pitch variations $X(t)$ as a function of time
          for the first part of the {\it Ricercar}. Pitch unit is the
          half-tone and the value $60$ corresponds to the keyboard midrange
          C (C3 = 60,\, C3$\sharp$ = D3$\flat$ = 61,\, D3 = 62, ...);
          time unit (on the horizontal axis) is the beat as given by the
          measure.

\noindent (b) Phase portrait: three-dimensional phase space trajectory
          obtained from the time series of the three parts of the
          {\it Ricercar} (see e.g. $X(t)$ in Fig.1.a).

\noindent (c) Time correlations: Normalized autocorrelation function $C(t)$
          of part one of the {\it Ricercar} ($X(t)$, Fig.1.a);
          $C(t) = (1/T)\sum_{\tau =0}^{T} (X(\tau ) - <X>)(X(\tau + t) -
          <X>) \Big/ (X(\tau ) - <X>)^2$.

\noindent (d) Power spectrum: $log-log$ representation of $S(f)$, the
          spectral density of $X(t)$ (Fig.1.a), obtained by FFT of $C(t)$.
          Frequency unit is reciprocal time unit. The straight lines
          correspond to $1/f$ and $1/f^2$ spectra respectively.

\bigskip

\noindent FIGURE 2 : $Log-log$ plot of the dimensionality versus parametric
          entropy for {\it all} sequences analyzed. $D_f^R$ is the value
          computed from the reconstructed trajectory in three-dimensional
          phase space (by the time delay method). Different symbols are used
          for classical music ($\bullet$), jazz music ($\Box$), and computer
          generated music ($\Diamond$)(random sequences).

\bigskip

\noindent FIGURE 3 : Dimensionality presented in chronological order of
          the pieces. $D_f$ is the value obtained for each piece
          by averaging over the values $D_f^R$ of each part of each piece
          (unless single part piece, BPS).
          The acronyms are defined in Table 1.

\bigskip

\noindent FIGURE 4 : First order parametric entropy $H'_1$; same as
          for $D_f$ (see caption of Fig.3).

\bigskip

\noindent FIGURE 5 : $Log-log$ representation of $D_f$ (see Fig.3) as
          a function of $H'_1$ (see Fig.4). Symbols are : $\bullet$
          for classical music, $\Box$ for jazz music, $\Diamond$ for
          random sequences.


\begin{references}
\bibitem{fo0} Dynamics is to be understood in the sense of the physicist
 as the (set of) process(es) governing  the time evolution of the system
 considered and should not be confused with the definition commonly used
 by musicians for loudness variations.
\bibitem{fo1} Intensity variations (loudness) and timbre are not
 considered here.
\bibitem{fo2} For instance conjecturing a power law $D_f \propto (H'_1)^\xi $,
 we found that a fit to the data produced a correlation coefficient whose
 value rules out the quantitative validity of the power law as representative
 of a possible relation between dimensionality and entropy. From Fig.5 it
 can be guessed that a third degree polynomial would yield an acceptable
 fit for which we found indeed a correlation coefficient of $.75$; however
 the interpretation of a cubic function could hardly be justified.
\bibitem{fo3} Note that white noise algorithm data redistributed independently
 over both pitch and time variables produce long time correlations and
 consequently the power spectrum of the corresponding sequence is not flat.
 Sequences so generated may qualify as examples of random music but should
 not be considered as white noise music if the criterion is the flatness of
 the spectrum. This point, which has been overlooked by some authors
 \cite {Vos}, was discussed by Nettheim \cite {Net}.
\end{references}
\end{document}